\documentclass[conference]{IEEEtran}

\usepackage{ifpdf}
\usepackage{cite}
\usepackage{amsmath,amssymb,amsfonts}
\usepackage{algorithmic}
\usepackage{graphicx, adjustbox}
\usepackage{array}
\usepackage{enumitem}
\usepackage{breqn}
\usepackage{ dsfont }
\usepackage{ upgreek }
\usepackage{textcomp}
\usepackage{multirow}
\usepackage{algorithm}
\usepackage{algorithmic}
\usepackage{pgfplots}
\usepackage{tkz-euclide}
\usepackage{subcaption}
\usepackage{tikz}{\tiny }
\usepackage{hyperref}
\usepackage{balance}
\usepackage{caption}
\usepackage{lipsum}

\begin{document}

\title{A Cooperative Architecture of Data Offloading and Sharing for Smart Healthcare with Blockchain
}
\author{\IEEEauthorblockN{Dinh C. Nguyen\IEEEauthorrefmark{1},
		 Pubudu N. Pathirana\IEEEauthorrefmark{1}, Ming Ding\IEEEauthorrefmark{2}, Aruna Seneviratne\IEEEauthorrefmark{3}
	}
	\IEEEauthorblockA{\IEEEauthorrefmark{1}School of Engineering, Deakin University, Australia \\
		\IEEEauthorrefmark{2}Data61, CSIRO, Australia  \\
		\IEEEauthorrefmark{3}School of Electrical Engineering and Telecommunications, UNSW, Australia
		\\
		Emails: \{cdnguyen, pubudu.pathirana\}@deakin.edu.au, ming.ding@data61.csiro.au, a.seneviratne@unsw.edu.au.
}
}
\maketitle

\begin{abstract}
The healthcare industry has witnessed significant transformations in e-health services where Electronic Health Records (EHRs) are transferred to mobile edge clouds to facilitate healthcare. Many edge cloud-based system designs have been proposed, but some technical challenges still remain, such as low quality of services (QoS), data privacy and system security due to centralized healthcare architectures. In this paper, we propose a novel hybrid approach of data offloading and data sharing for healthcare using edge cloud and blockchain. First, an efficient data offloading scheme is proposed where IoT health data can be offloaded to nearby edge servers for data processing with privacy awareness. Then, a data sharing scheme is integrated to enable data exchange among healthcare users via blockchain. Particularly, a trustworthy access control mechanism is developed using smart contracts for access authentication to achieve secure EHRs sharing. Implementation results from extensive real-world experiments show the superior advantages of the proposal over the existing schemes in terms of improved QoS, enhanced data privacy and security, and low smart contract costs. 
\end{abstract}
\section{Introduction}

The recent advances of cloud computing, edge computing and Internet of Things (IoT) technologies, have empowered e-health services [1], [2]. In modern e-healthcare, health data collected from mobile devices-MD (i.e. smartphones and wearable sensors) can be offloaded to mobile edge computing (MEC) servers for efficient computation and analysis, then improving high quality of services (QoS) and reducing resource burden on devices. Particularly, cloud with resourceful servers can also be integrated to store historic health data analysed from the offloading phase, which enables data sharing among health users. For example, a doctor can exploit cloud data to support disease diagnosis, and patients can gain medical benefits like health or medication advice. The cooperation of data offloading and data sharing thus facilitates the delivery of health care services [3]. 

However, realizing the promises of such a cooperative system still faces non-trivial challenges. First, how to offload IoT healthcare data to edge-cloud for supporting efficiently health applications while guaranteeing both high QoS and data privacy is a critical issue. Most of traditional approaches [4-6] only either focus on the QoS problem of network latency and energy usage or data privacy for the healthcare offloading, while implementing a holistic framework with all these factors taken into consideration is vitally necessary. Second, the centralized cloud architectures remain single-point failures which potentially disrupts the entire network [7]. Moreover, the EHRs storage on central cloud adds communication overhead for data retrieval, although it requires less data management efforts. Third, it is not straightforward to implement secure data sharing in e-health networks where there is often a lack of transparency and trust among participants [2]. Attackers or curious users can access health data without users' permission, leading to leakage risks of sensitive patient information. Final, the feasibility and implementation of such a hybrid approach of data offloading and data sharing for healthcare applications remain unsolved in most existing works [7-13], [15], which urgently requires further innovative solutions. 

To overcome the above challenges, this paper presents a novel cooperative architecture of data offloading and data sharing for healthcare using edge-cloud and blockchain. Edge computing is employed to offer cost-efficient offloading services for improving QoS, while privacy in computation is ensured by data encryption. We develop a decentralized storage system on cloud and employ smart contracts for reliable data sharing so that system latency and security requirements can be met. Due to resource constraints, MDs are regarded as lightweight nodes and only participate in the blockchain network for data collection or sharing, while mining works are done by resourceful cloud machines. The main purpose of blockchain adoption is to use its decentralization and security for building a distributed cloud system and a secure data sharing scheme, which effectively solve high latency overhead and single-point failure issues faced by conventional architectures [4-9]. We also conduct extensive real-world experiments to verify the feasibility of the proposed joint framework. In a nutshell, this article provides a set of contributions as follows:
\begin{enumerate}
	\item We first propose an efficient data offloading scheme where IoT health data can be offloaded to nearby edge servers for data processing with privacy awareness.
	\item	We then propose a new data sharing scheme which is integrated to enable data exchange among healthcare users via blockchain. A trustworthy access control mechanism is also developed using smart contracts for access authentication to achieve secure EHRs sharing.
	\item	We conduct various experiments to verify the advantages of the proposed approach over other baseline methods in terms of offloading and sharing performances. 
\end{enumerate}

The remainder of the paper is organized as follows. Section II discusses related works. We propose an integrated architecture in Section III with offloading and sharing formulation. Section IV presents implementation results on various performance metrics, and Section V concludes the paper. 
\vspace{-0.06in}
\section{	Related Works}
In this section, we survey the related works in data offloading and data sharing for healthcare. 
\vspace{-0.07in}
\subsection{Health Data Offloading}
Many data offloading approaches have been proposed to support healthcare. In [4], mobile healthcare data can be offloaded to fog nodes or cloud for processing, analysis, and storage. In [6], [7], a multi-cloud model was proposed which enables offloading of mobile health data to the cloud under latency and energy constraints. The main drawback of such proposals is the high latency incured by offloading data to remote clouds. Also, offloading privacy is not considered, which puts sensitive health data at risks of external attacks. Another work in [8] proposed an IoT architecture for executing healthcare applications on clouds, but optimization for memory usage of MDs required to offload the data and data privacy concerns are completely neglected. Meanwhile, other works [5], [9], [10] concentrated on offloading security issues in healthcare. For example, [9] used hash function and key cryptosystem for data security. Also, privacy issues for health data offloading were also solved in [5], [10] by using consensus algorithms and learning-based privacy preservation techniques with respect to response time and delay. However, the above studies lack the joint consideration of all QoS constraints (network latency, energy consumption and memory usage) and privacy awareness, which is of significant importance for offloading performance guarantees [2]. 
\vspace{-0.06in}
\subsection{Health Data Sharing}
Several solutions using blockchain are proposed for health data sharing. The work [11] introduced a privacy-preserved data sharing scheme enabled by the conjunction of a tamperproof consortium blockchain and cloud storage. Furthermore, [12] described a hybrid architecture of using both blockchain and edge-cloud nodes where smart contracts are employed to monitor access behaviours and transactions. Despite data privacy enhancements, such solutions [11], [12] mainly rely on central cloud servers for EHRs storage, which remains single-point failure bottlenecks and incurs high communication overhead. Further, the performances of smart health contract have not been evaluated. Meanwhile, [13] employed an interplanetary file system (IPFS) with Ethereum blockchain for EHRs sharing over clouds, but data retrieval speed and security capability, which are important performance metrics, have not been verified. Recently, our works [3], [14] showed a first attempt to implement a mobile cloud EHRs sharing using decentralized storage IPFS and smart contract. The study in [15] investigated an access control protocol based on blockchain and external public key infrastructure (PKI), but it requires complex and expensive resources to achieve secure EHRs sharing [2]. 

Despite promising results, the aforementioned works fail to provide a cooperative framework of data offloading and data sharing for healthcare. This motivates us to develop a comprehensive solution by leveraging MEC and blockchain to provide better healthcare services in terms of enhanced QoS, improved privacy and security. 
\vspace{-0.06in}
\section{	Proposed Architecture and System Design}

\begin{figure*}
	\centering
	\includegraphics[width=0.99\linewidth]{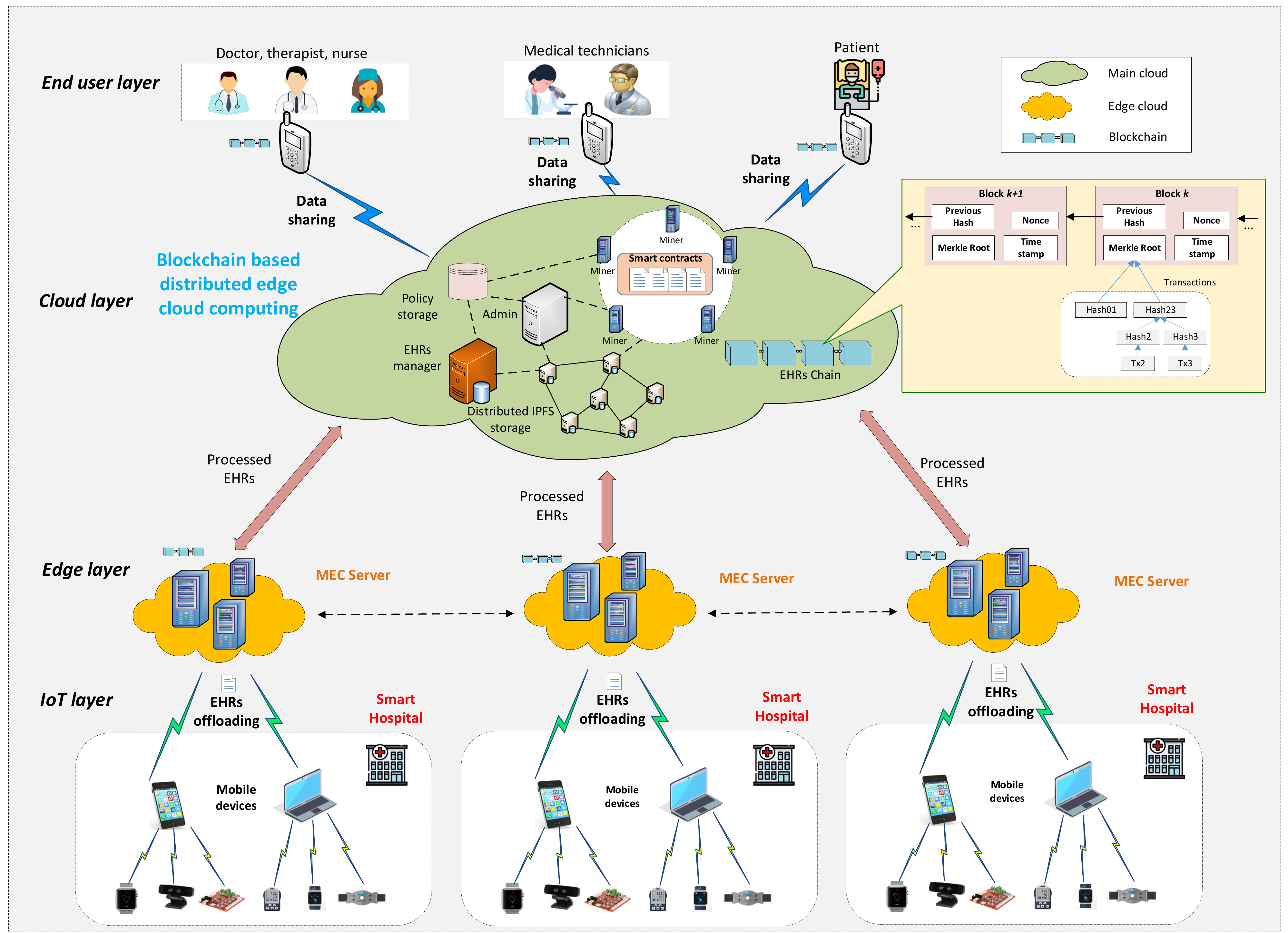} 
	\caption{The proposed healthcare architecture. }
	\label{Fig:Overall}
\end{figure*}

We consider a healthcare system architecture in Fig.~\ref{Fig:Overall}, consisting of four layers as follows. (1) \textit{IoT layer} consists of many smart hospitals which monitor patients by MDs as mobile gateways in different locations from sensor IoT devices. (2) \textit{Edge layer} includes a cluster of edge cloud nodes, each edge node manages a group of nearby IoT devices to provide distributed computing services for healthcare. All computations including data processing and analysis are implemented at the edge layer to offer instant healthcare services. (3) \textit{Cloud layer} which stores processed health data from edge nodes and performs data sharing with end users. To build a cloud blockchain network, we create four key cloud components, including admin, EHRs manager, distributed cloud storage  and smart contracts with miners and policy storage. Their details will be described in the next section. (4) \textit{End user layer} is the network of healthcare users such as healthcare providers, caregivers and patients, who are interested in using cloud healthcare services. For instance, doctors use analysed health data on cloud for disease diagnosis, or patients can track their medical record history. Note that the blockchain network here consists of edge servers, cloud entities and end users, and are maintained by secure transaction ledgers distributed over the blockchain participants [14]. Next, we focus on the analysis of data offloading and data sharing schemes.

\subsection{Health Data Offloading}
 We consider that each MD has multiple health data tasks  $\mathcal{N} = \{1,2,..., N\}$ to be executed. We introduce an offloading decision policy denoted by a binary variable $x_{n}^t \in \{0,1\} $, where $x_{n}^t = 1$ means that task $n$ is offloaded to the edge server, otherwise it is executed locally $x_{n}^t = 0$. 
\subsubsection{Offloading Model}
\begin{figure}
	\centering
	\includegraphics[width=0.95\linewidth]{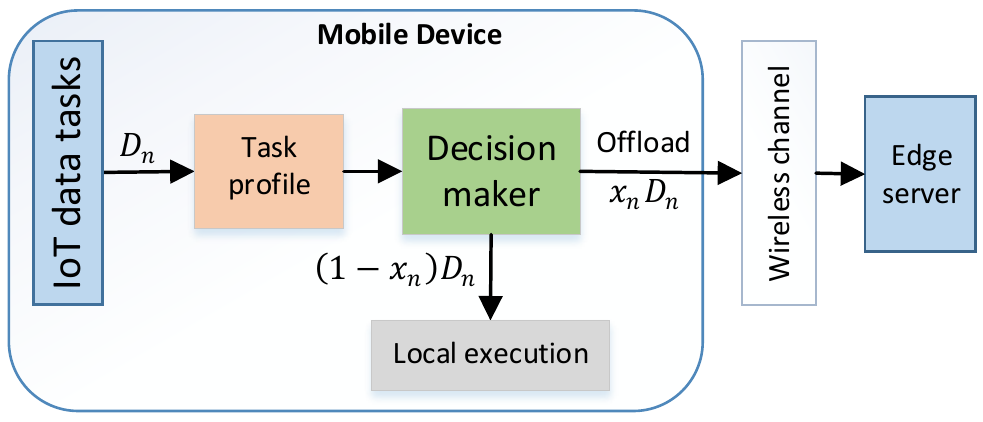}
	\caption{The data offloading scheme.  }
	\label{Fig:offloading}
\end{figure}

Motivated by experimental results of our recent work [16], in this paper we propose an offloading architecture as Fig.~\ref{Fig:offloading} which includes two main modules: task profile and decision maker on mobile devices.
 
- \textit{Task profile}: This module collects device information such as energy consumption ($E$), processing time ($T$) and memory usage ($M$) when executing data tasks, by using mobile performance measurement tools. Therefore, a task profile with a size $D_n$ (in bits) can be formulated as a variable tuple $[D_n, E_n, T_n, M_n]$ which is then stored in a database created on the MD for supporting offloading decisions. 

- \textit{Decision maker}: This module receives task profile information collected by the profile module to make offloading decisions. Similar to [17], [18], we employ an integer linear programming model to develop a decision making algorithm on MDs. By using profile information, the algorithm analyses and makes decisions for executing locally or offloading to the MEC server. The main objective is to determine an optimal computation decision for each task to minimize computing latency, energy consumption and memory usage. 

\subsubsection{	Offloading Formulation}
Motivated by healthcare offloading studies in [10] and [18], we formulate the health data offloading problem with three main metrics, namely processing time, energy consumption and memory usage under two computation modes.

- \textit{Local execution}: When a MD decides to execute the task \textit{n} locally ($x_{n} =0$), it uses its resource to process healthcare data. We denote $X^l_n,f^l_n$ as mobile CPU utilization for task $n$ (in CPU/bit) and mobile computational capacity (in CPU/sec), respectively. Then, the local execution time can be calculated as $\scriptstyle T^{local}_n = \dfrac{D_nX^l_n}{f^l_n}$. We also define $E^{local}_n$ and $M^{local}_n$ as battery consumption (in Mah) and memory usage (Mbyte), which can be measured by mobile measurement tools [18]. 

- \textit{Offloading to MEC server}: In the case of task offloading ($x_{n} =1$), the data task needs to be encrypted for security before transmitting to the MEC. We denote $X^{enc}_n, X^e_n,f^e_n$ as mobile CPU utilization for encrypting the task $n$ (in CPU/bit), edge CPU utilization (in CPU/bit), and edge computational capacity (in CPU/sec). Further, let denote $r_n$ as the transmission data rate of the MD, the total offloading time can be expressed as $\scriptstyle T^{offload}_n = (\dfrac{D_nX^{enc}_n}{f^l_n} + \dfrac{D_nX^{e}_n}{f^e_n} +\dfrac{D_n}{r_n})$. We also define $E^{enc}_n, E^{trans}_n$ as encryption energy and energy for transmitting the task $n$ to the MEC. Then, the total offloading energy is computed by [16] $\scriptstyle E^{offload}_n =(E^{enc}_n\dfrac{D_nX^{enc}_n}{f^l_n} +E^{trans}_n\dfrac{D_n}{r_n}) $. Moreover, the offloading process also incurs a memory usage cost for encryption, defined as $M^{offload}_n$, which can be also obtained through mobile measurement tools [18].

Accordingly, the total offloading time, energy cost and memory usage can be expressed as follows.
\begin{equation}
 T_n = (1-x_n)T^{local}_n + x_nT^{offload}_n,
\end{equation}
\begin{equation}
 E_n = (1-x_n)E^{local}_n + x_nE^{offload}_n,
\end{equation}
\begin{equation}
 M_n = (1-x_n)M^{local}_n + x_nM^{offload}_n.
\end{equation}

Based on above formulations, we can derive the optimization problem to jointly optimize time latency, energy cost and memory usage under system constraints as follows

\begin{equation}
\begin{aligned}
&  \underset{\textbf{x}}{\text{min}}
&& \sum_{n=1}^{N} (\alpha_tT_n+\alpha_eE_n+\alpha_mM_n)\\
& \text{st.}
&& (C1): \sum_{n=1}^{N} (x_nT^{offload}_n) \leq \sum_{n=1}^{N} (1-x_n)T^{local}_n,\\
&&& (C2): \sum_{n=1}^{N} (x_nE^{offload}_n) \leq \sum_{n=1}^{N} (1-x_n)E^{local}_n),\\
&&&(C3): (\sum_{n=1}^{N} (x_nT^{offload}_n) + \sum_{n=1}^{N} (1-x_n)T^{local}_n) \leq \tau, \\
&&& (C4): (\sum_{n=1}^{N} (x_nM^{offload}_n) + \sum_{n=1}^{N} (1-x_n)T^{local}_n) \leq \zeta,
\end{aligned}
\end{equation}
where $\alpha_t, \alpha_e, \alpha_m$ are the cost weights and all set to 1/3, respectively. Here, the constraints (C1), (C2) represent that the offloading cost of time delay and energy consumption should be less than the local execution cost when computing all healthcare data tasks on a MD. In fact, the solution of offloading tasks to the MEC should be preferred due to its efficient computation, especially for large-size tasks for better QoS. Further, the total task execution time should not exceed a maximum latency value (C3). Meanwhile, (C4) defines that the memory used for task computation must not exceed the available mobile memory. In this paper, we employed the particle swarm optimization (PSO)\footnote{ https://github.com/topics/particle-swarm-optimization} model written in java to build the above offloading optimization algorithm on Android phones. The PSO algorithm has proven its superior advantages over its counterparts like Genetic Algorithm (GA) regarding extremely low computational cost and simple implementation on Android devices for mobile offloading applications like healthcare [19]. 
\subsection{Health Data Sharing}
We describe main components, smart contract design and access protocol for data sharing as follows. 
\subsubsection{Main Components}
In the data sharing scheme, we assume that healthcare data were processed and stored on cloud by the offloading framework as designed in the previous subsection. The combination of cloud computing and blockchain can enable highly efficient data sharing with improved throughput, reduced data retrieval and better security [2]. Note that blockchain utilizes consensus validation [2] to achieve synchronization on distributed ledgers which are replicated across cloud entities, avoiding single-point failures, i.e. the disruption of an entity does not impact the operation of the cloud system thanks to strongly linked transactions over the blockchain [2]. We propose a sharing architecture on cloud as shown in Fig.~\ref{Fig:Overall}, consisting of four main entities as follows. 

- \textit{EHRs manager: } It is responsible to control all user transactions on the blockchain network, including data storage processes of MDs and data access of mobile users. The management capability of EHRs manager is enabled by smart contracts through strict user policies. 

- \textit{Admin:} It manages transactions and operations on cloud by adding, changing or revoking access permissions. Admin deploys smart contracts and the only entity with the ability to update or modify policies in smart contracts. 

- \textit{Smart contract: } It defines all transactions allowed in the access control system. Users can interact with smart contracts by the contract address and Application Binary Interface (ABI). Smart contracts can identify, validate access and grant permissions for healthcare data request. It is considered as core software in our healthcare platform. 

- \textit{Decentralized IPFS storage: }We deploy on cloud a decentralized peer-to-peer file system InterPlanetary File System (IPFS), a network of distributed storage nodes to build a storage and sharing platform in the blockchain network [14]. Health results which were analysed from the offloading scheme are stored in identified IPFS nodes, while their hash values are recorded by EHRs manager and stored in Distributed Hash Table (DHT). We also integrate smart contracts with IPFS to improve decentralized cloud storage and controlled data sharing for better user access management. Details of IPFS settings can be seen in our recent work [14]. 

\subsubsection{	Smart Contract Design}

We first create a sharing contract controlled by the admin to monitor transaction operations. We denote $PK$ as the user's public key, $userRole$ as the user's role, $Addr$ as the patient's address in blockchain. The contract mainly provides the following five functions.
\begin{itemize}
	
	\item \textit{AddUser(PK, userRole):} (executed by Admin) This fucntion allows to add a new user to the main contract. User is identified by their public key and is added into the contract with a corresponding role based on their request.
	\item \textit{DeleteUser(PK, userRole):} (executed by Admin) It is used to remove users from the network based on the corresponding public key. All personal information is also deleted from cloud storage.
	\item \textit{PolicyList(PK):} (executed by Admin) A peer of health provider-patient can agree on a policy which expresses their healthcare relation. For example, a patient has an identified doctor for his health care and only this doctor has rights to access EHRs of his patient. The policy list contains users' public keys in policy storage for identification when the smart contract processes new transactions.   
	\item \textit{RetrieveEHRs(PK, Addr):} (executed by EHRs manager) It allows to retrieve cloud EHRs of patients. A blockchain entity needs to provide the address of patient (including \textit{Patient ID} and \textit{Area ID}) to the smart contract. The contract then verifies and sends a message to the EHRs manager to extract and return data to the requester. 
	\item \textit{Penalty (PK, action):} (executed by Admin) When detecting an unauthorized request to EHRs system, the EHRs manager will inform smart contract to issue a penalty to the requester. In our paper, we give a warning message as a penalty to the unauthorized mobile entity. 
\end{itemize} 
Next, we present a data sharing protocol as summarized in the following four steps. 
\\
\textit{\textbf{Step 1: Request processing (executed by EHRs manager)}}
\\
The EHRs manager receives a new request as a transaction \textit{Tx} associated with request IDs (including \textit{AreaID} and \textit{PatientID}) from a mobile user (i.e. a health provider or a patient). The EHRs manager will obtain the requester's PK by using the \textit{Tx.getSenderPublicKey()} function and send it to the contract for validation. 
\\
\textit{\textbf{Step 2: Verification (executed by the Admin)}}
\\
After receiving a transaction from EHRs manager $(msg.sender = ME)$, the admin will verify the request based on its $PK$ in the policy list of the smart contract. If the $PK$ is available in the list, the request is accepted and now a data access permission is granted to the requester. Otherwise, the smart contract will issue a penalty through the \textit{penalty()} function to discard this request from the blockchain network. 
\\
\textit{\textbf{Step 3: EHRs retrieval (executed by the Admin)}}
\\
Once the permission is granted, the contract will decode the transaction using the \textit{abiDecoder:decodeMethod(Tx)} function to obtain the address information of EHRs in the data field of transaction (see Section II). Now the admin can know the \textit{Area ID} and \textit{Patient ID} of the request, and then forward it to EHRs manager for data retrieval from IPFS [14]. 
\\
\textit{\textbf{Step 4: Data feedback (executed by EHRs manager)}}
Once the requested data is found, the EHRs manager will send it to the requester. Now the sharing is finished and a new transaction is appended to blockchain and broadcasted to network users. Note that data in such transactions are mainly patient addresses, which are lightweight and efficient to store on the blockchain. Algorithm 1 shows our sharing protocol, and its source code is available in our recent work [14].
\vspace{-0.07in}
\begin{algorithm}
	\footnotesize
	\caption{EHRs access protocol}
	\begin{algorithmic}[1]
		\STATE \textbf{Input:} $Tx$ (The data request on blockchain)
		\STATE \textbf{Output:} $Result$  (Access result)
		\STATE \textbf{Initialization:} \textit{(by the EHRs Manager)} 
		\STATE Receive a new transaction $Tx$ from an end user
		\STATE Obtain the PK: $PK\leftarrow Tx.getSenderPublicKey() $
		\STATE Send the public key to Admin (\textit{msg.sender = EHRs Manager})
		\STATE \textbf{Pre-processing the request} \textit{(by Admin)} 
		\IF {$PK$ is available in the policy list}
		\STATE  $PolicyList(PK) \leftarrow true$
		\ENDIF
		\STATE $decodedTx \leftarrow abiDecoder.decodeMethod(Tx) $
		\STATE  $Addr \leftarrow web3.eth.getData(decodedTx([DataIndex])$
		\STATE Specify \textit{DeviceID}: $D_{ID} \leftarrow Addr(Index[D_{ID}])$; 
		\STATE \textbf{Verification} \textit{(by the smart contract)} 
		\WHILE {true}
		\IF {$PolicyList(PK) \rightarrow true$}
		\IF {$PolicyList(D_{ID}) \rightarrow true$}
		\STATE $Result \leftarrow RetrieveEHRs(PK,Addr)$
		\STATE break;
		\ELSE
		\STATE $Result \leftarrow Penalty(PK; action)$
		\STATE break;
		\ENDIF
		\ELSE
		\STATE $Result \leftarrow Penalty(PK; action)$
		\STATE break;
		\ENDIF
		\ENDWHILE
	\end{algorithmic}
\end{algorithm}

\section{Experimental Results and Evaluations}
In this section, we present experiments and perform implementation evaluations in details. 
\subsection{Experiment Settings}
\begin{figure}
	\centering
	\includegraphics[width=0.95\linewidth]{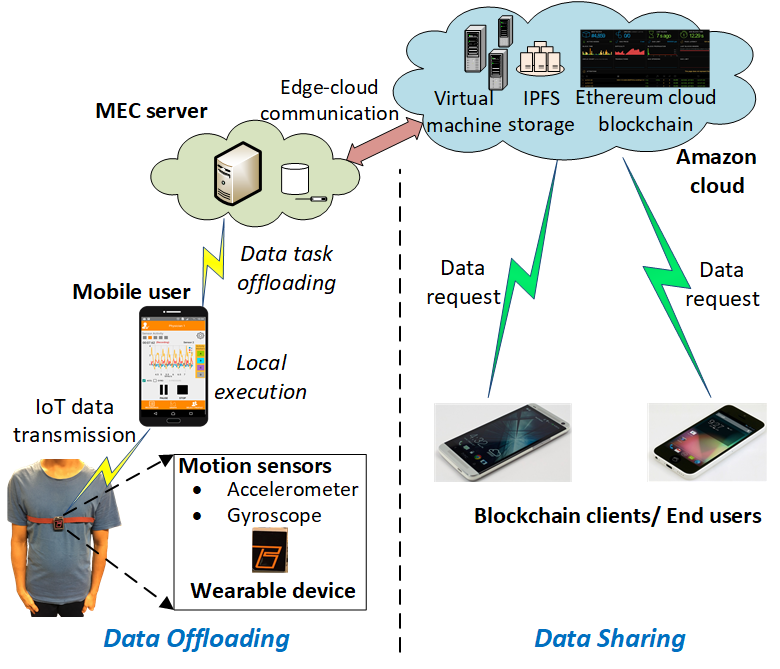}
	\caption{ Experiment setting. }
	\label{Fig:experiment}
\end{figure}
 We implemented a full experiment with data offloading and data sharing tests to prove the proposed scheme, as shown in Fig.~\ref{Fig:experiment}. 
  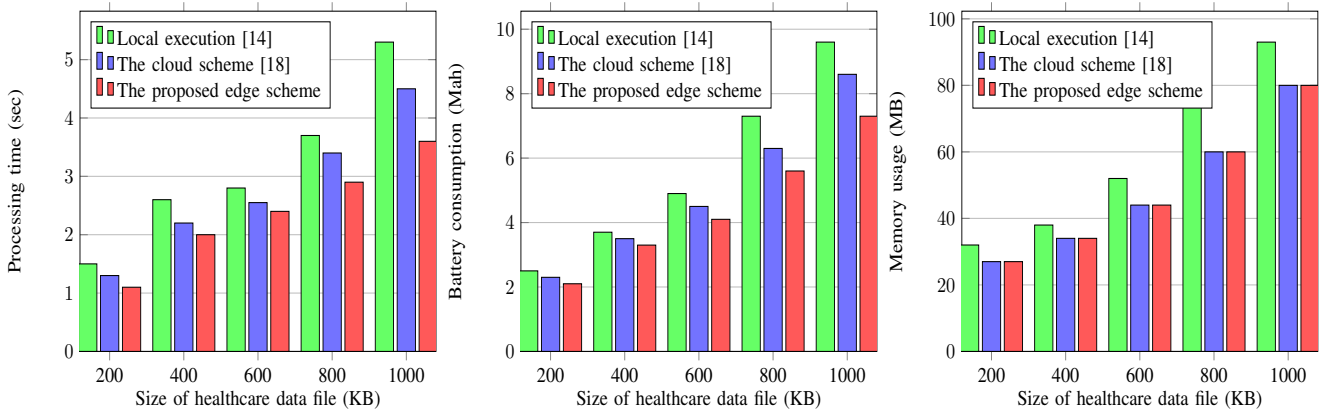
\begin{figure*}
 	\centering
 	\begin{adjustbox}{ height=5.5cm, width= 17.5cm}
 		\begin{tikzpicture}
 		\begin{axis}[ybar,legend pos=north west, legend cell align={left},
 		ylabel=Processing time (sec),ymajorgrids,
 		xlabel = Size of healthcare data file (KB),
 		symbolic x coords={200, 400, 600, 800, 1000},
 		xtick=data,ymin=0
 		]
 		\addplot [fill=green!60] coordinates {(200,1.5) (400,2.6)   (600,2.8) (800,3.7) (1000,5.3)   };
 		
 		\addplot [fill=blue!55] coordinates {(200,1.3) (400,2.2)   (600,2.55) (800,3.4) (1000,4.5)   };
 		
 		\addplot [fill=red!65] coordinates {(200,1.1)  (400,2)  (600,2.4)  (800,2.9)  (1000,3.6) };
 		
 		\legend{Local execution [14], The cloud scheme [18], The proposed edge scheme }
 		\end{axis}
 		\end{tikzpicture}
 		
 		\begin{tikzpicture}
 		\begin{axis}[ybar,legend pos=north west, legend cell align={left},
 		ylabel=Battery consumption (Mah),ymajorgrids,
 		xlabel = Size of healthcare data file (KB),
 		symbolic x coords={200, 400, 600, 800, 1000},
 		xtick=data,ymin=0
 		]
 		\addplot [fill=green!60] coordinates  {(200,2.5)  (400,3.7)  (600,4.9)  (800,7.3)  (1000,9.6) };
 		
 		\addplot [fill=blue!55] coordinates  {(200,2.3)  (400,3.5)  (600,4.5)  (800,6.3)  (1000,8.6) };
 		\addplot [fill=red!65] coordinates  {(200,2.1)  (400,3.3)  (600,4.1)  (800,5.6)  (1000,7.3)  };
 		
 		\legend{Local execution [14], The cloud scheme [18], The proposed edge scheme}
 		\end{axis}
 		\end{tikzpicture}
 		
 		\begin{tikzpicture}
 		\begin{axis}[ybar,legend pos=north west, legend cell align={left},
 		ylabel=Memory usage (MB),ymajorgrids,
 		xlabel = Size of healthcare data file (KB),
 		symbolic x coords={200, 400, 600, 800, 1000},
 		xtick=data,ymin=0
 		]
 		\addplot [fill=green!60] coordinates  {(200, 32)  (400,38)  (600,52)  (800,77)  (1000,93)  };
 		
 		\addplot [fill=blue!55] coordinates  {(200,27)  (400,34)  (600,44)  (800,60)  (1000,80)  };
 		
 		\addplot [fill=red!65] coordinates  {(200,27)  (400,34)  (600,44)  (800,60)  (1000,80)  };
 		
 		\legend{Local execution [14], The cloud scheme [18], The proposed edge scheme}
 		\end{axis}
 		\end{tikzpicture}
 	\end{adjustbox}
 	
 	\caption{Experimental results for local, cloud and edge computation.} 
 	\label{Fig:Experimentalresults}
 	\vspace{-0.1in}
 \end{figure*}

 For the health data offloading implementation, we employed the Lambda Edge [20] service enabled by an Amazon EC2 server (Intel Xeon Family), CPU 2.5 GHz, 2 GB memory and maximum network bandwidth 3500 Mbps. We used a Sony Android mobile phone as a MD with Qualcomm Snapdragon 845 processor, 1GB memory, and a battery capacity of 2870mAh. The MD connects with the edge cloud computing on the wireless network via Wi-Fi with a maximum data rate of 11 Mbit/s. For data encryption, we used a symmetric algorithm AES to preserve data due to its less time and energy consumption [18] which is well suitable for low-latency health applications.
 
 Healthcare data and programming code are necessary for our test. For a specific use case, we used Biokin sensors [3] as IoT devices to collect simultaneously human motion data (acceleration and gyroscope time-series data) and store in separate files to be executed by both MDs and the edge server. By using our data analysis algorithm, we can specify human movement severity levels (i.e. movement disorders) to serve doctors during clinical decisions [3]. For mobile performance evaluations, we employed Firebase Performance Monitoring service [21] to measure processing time, battery consumption, and memory usage. The mobile application for offloading optimization mentioned in Section III.A was implemented using Android studio 3.5. Meanwhile, for the evaluation of edge execution, we utilized the Kinesis Data Analytics service available on Amazon cloud to monitor data streaming and measure computation. 
 
 For data sharing experiment, we deployed a private Ethereum blockchain network supported by Amazon cloud where two virtual machines AWS EC2 were employed as the miners, two virtual machines Ubuntu 16.04 LTS were used as the admin and EHRs manager, respectively. The decentralized IPFS storage was integrated with Amazon cloud and its network configuration was presented in [14]. Our smart contract was written by Solidity programming language and deployed on AWS Lambda functions and its source code is available in [14]. Users can interact with smart contracts through their Android phone where a Geth client was installed to transform each smartphone into an Ethereum node. We also used two Android phones to investigate sharing results. More details of hardware configurations and parameter settings for our system are described in our recent works [3], [14], [16].
 
 \subsection{Experiment Results}
 \subsubsection{Data Offloading Performance}
We compare our scheme with two baselines: local execution [14] (only executing data on devices) and cloud computation [18] (offloading to the cloud server) to prove the advantages of our scheme. A set of health data files with different sizes (200 KB-1200 KB) [18] collected from sensors was used in evaluations. We implement each test with 10 times to obtain average values, and evaluate via three performance metrics: processing time, energy consumption, and memory usage as shown in Fig.~\ref{Fig:Experimentalresults}. 

For the processing time, it consists of execution time for the local case and encryption time, offloading time and remote execution time for the cloud and edge case. Based on results in Fig.~\ref{Fig:Experimentalresults}, the proposed edge scheme achieves the best performance in terms of the average processing time. For example, executing a 200 KB file by the edge scheme only consumes 1.1 sec, whereas it reaches about 1.3 sec and 1.5 sec in the cloud and local schemes, respectively. This leads to a 10-18\% time saving of data execution by using edge computing. Further, the proposed edge scheme saves up to 31\% and 15\% time when computing a 1200 KB file, compared to local and cloud schemes, respectively. We also found with the selected human motion dataset, although data encryption is integrated in offloading, the edge-cloud offloading schemes still achieves better offloading performances than the local scheme, showing the efficiency of the proposed encryption technique. 

For battery consumption, health data tasks consume less energy when being executed with the edge offloading scheme. As an example, offloading a 200 KB file consumes less 11\% energy than the case of local computation and less 5\% energy than the cloud scheme. Specially, the energy usage of the edge scheme becomes more efficient when the data size increases. For instance, executing a 1000 KB and 1200 KB file can save 21.3\% and 28.1\% energy, respectively when offloading the task to the edge server, while the cloud and local schemes consume higher energy. For memory performance, the edge and cloud schemes has the same memory usage due to using the same encryption mechanism for security. However, these schemes achieve greater memory performances, with 5\% and 9\% memory savings compared to the local scheme when executing a 200 KB and 1200 KB file, respectively. Note that the above implementation results were obtained from the proposed offloading application with human motion data and current hardware settings of devices and edge servers. Different mobile applications with other health data types such as Electroencephalography (EEG) or video data and different hardware settings can achieve different offloading performances [7]. However, generally the proposed edge offloading scheme yields the best performances with enhanced time latency, energy, and memory usage and shows superior advantages than the cloud scheme and local scheme when the size of health data increases.

\begin{figure*}
	\centering
	\includegraphics[width=0.95\linewidth]{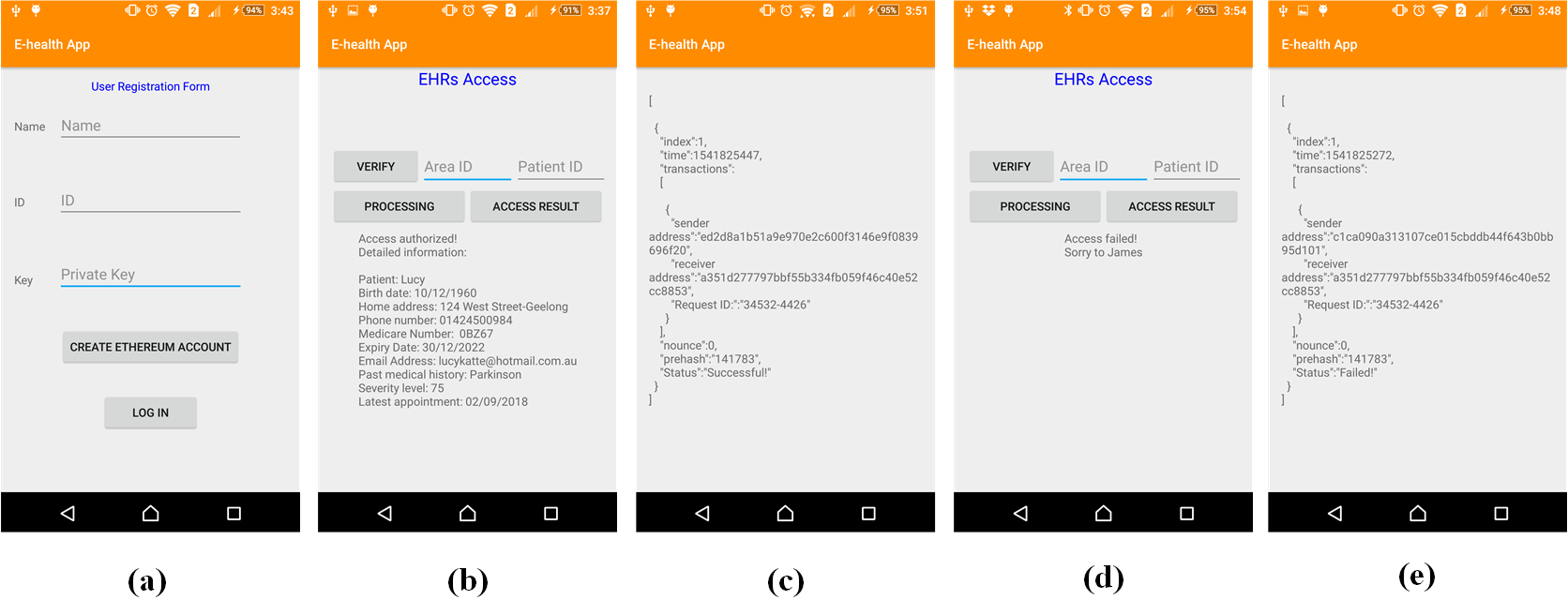}
	\caption{Data sharing results on Android phones: a) User Registration form with Ethereum account, b) Data access results of an authorized user, c) Transaction record of authorized access, d) Access result of an unauthorized user, e) Transaction record of unauthorized access.}
	\label{Fig:Datasharing}
\end{figure*}

\subsubsection{Data Sharing Performance} We investigated  two main performance metrics: access control and network overheads for the proposed data sharing.

We present two use cases with authorized and unauthorized access to evaluate access control, as shown in Fig.~\ref{Fig:Datasharing}. The goal is to enable end users to retrieve effectively EHRs on cloud and prevent malicious access to our cloud data. A mobile user, i.e a doctor, who wants to access EHRs of his patient on cloud, can use an Ethereum account to register user information for joining the blockchain (Fig.~\ref{Fig:Datasharing}(a)). After his request is verified by the cloud EHRs manager, he now starts to make a transaction to access EHRs by providing the address of his patient (including \textit{AreaID} and \textit{PatientID} as defined in the offloading scheme) as shown in Fig.~\ref{Fig:Datasharing}(b). Our sharing system will then return data access results which are also updated on his mobile interface (Fig.~\ref{Fig:Datasharing}(b)). Thus, the doctor can acquire patient's disability levels (scores) computed from the offloading phase for medications. Once the EHRs access process is finished, a new transaction is appended into blockchain by mining process and broadcast to all participants. Patients thus can monitor sharing transactions and know who uses their data (Fig.~\ref{Fig:Datasharing}(c)), thus ensuring user data ownership and network trustworthiness.

In the case of unauthorized access, the smart contract will verify and detect by the access protocol with a predefined policy list. Such illegal request is prevented and discarded from our EHRs database, and a warning message is returned to the requester (see Fig.~\ref{Fig:Datasharing}(d)). A corresponding transaction for unauthorized access is also issued by the smart contract (see Fig.~\ref{Fig:Datasharing}(e)). Obviously, blockchain is capable of controlling data access and thus improve system reliability and data privacy. Further security analysis is presented in the next sub-section. 

\begin{table}
	\scriptsize
	\centering
	\caption{{Comparison results for data retrieval speeds (in sec)}}
	\label{table1}
	
	\begin{tabular}{|c||c c c c c c|}
		\hline
		\multirow{2}{*}{\textbf{Schemes}} &
		\multicolumn{6}{c|}{\textbf{Number of mobile user}} \\
		& \textit{N=2} & \textit{N=4} & \textit{N=6} & \textit{N=8} & \textit{N=10} & \textit{N=12}  \\
		\hline
		Centralized storage [11] & 1.6 & 2.4& 3.9 & 4.8 & 5.5 & 7.8 \\
		
		Proposed IPFS storage& \textbf{0.6} & \textbf{1.6}& \textbf{2.6} & \textbf{3.5} & \textbf{4.4} & \textbf{5.3} \\
		\hline
	\end{tabular}
	
\end{table}

\begin{table}
	\scriptsize
	\centering
	\caption{{Smart contract cost test}}
	\label{table2}
	
	\begin{tabular}{|c||c c c|}
		\hline
		\textbf{Contract functions}  & \textbf{Gas used} & \textbf{Actual cost (ether)} & \textbf{USD} \\
		\hline
		AddUser & 34603&	0.00069	&0.1168239 \\
		
		DeleteUser &	12098&	0.00024	&0.0406344\\
		PocicyList &	90684 &	0.0018 &	0.304758 \\
		
		RetrieveEHRs &	862409	 &0.0172 &	2.912132\\
		
		Penalty	 &573783 &	0.01147	 &1.9419857  \\
		\textbf{Total} & \textbf{1573577} & \textbf{0.0314}& \textbf{5.316334}\\
		\hline
	\end{tabular}

\end{table}

Furthermore, we also investigated time overhead of data sharing as shown in Table~\ref{table1}. Multiple MDs can access simultaneously the IPFS storage for data retrieval and time latency is measured. Specially, we compared our design with decentralized IPFS storage with the baseline [11] which utilized the conventional central cloud storage for sharing. The experiment results clearly show that the proposed decentralized storage scheme on IPFS cloud blockchain has significantly less time overhead as compared to the conventional scheme with centralized storage. For example, the proposed scheme can save 17\% and 30\% time for retrieving data on cloud in the case of 6 users and 12 users, respectively, in comparison with the baseline, which shows a significant advantage of the proposed IPFS-based storage approach.

\subsection{Smart Contract Performance}

To evaluate the performance of the smart contracts in our healthcare system, we investigated the operation costs of contract functions when there are 5 mobile users on our Ethereum blockchain as listed in Table~\ref{table2}. The cost is calculated in gas unit and then converted into ether (cost unit of Etherum blockchain) and US dollars by using an exchange rate of 1 Gas $\approx$ 0.00000002 Ether and 1 Ether $\approx$ \$169.31 at the time of this study. We consider a realistic scenario that some new users can join the healthcare network, some current users can leave, and therefore $AddUser$ and $DeleteUser$ functions need to be executed. Furthermore, the contract can also allow data retrieval with $RetrieveEHRs$ for authorized users or force penalties with $Penalty$ for unauthorized users. All of these contract executions incur operation costs and the users need to pay for their service usage. From Table~\ref{table2}, the amount of gas used for sharing services is 1573577 gas (5.316334 USD, $\approx$ 1.063 USD per user). Clearly, the financial cost for using our contract is low, which demonstrates the practicality of the proposed contract-based data sharing scheme.

\subsection{Attack Models and Security Analysis}
We consider two potential threat types as follows. External threats: during data offloading and sharing, external attackers can gain access to obtain health information. Insider threats: network participants may be untrusted and retrieve EHRs without users' consent. Our design can address these issues and gain more security benefits than current works [7-12]. 

- We employ an AES encryption on MDs when performing offloading to encrypt healthcare data before transmitting to the edge server. This would establish a new security layer between devices and edger server to protect sensitive health information against external attack threats and thus improve data confidentiality. Furthermore, the proposed decentralized IPFS cloud system enables data storage on distributed virtual nodes on blockchain without central server. Once a data file of analysed health results is uploaded to the IPFS, its hash is automatically returned to the EHRs manager and this also updated in DHT table. Any modifications on data files in IPFS can be easily detected by the EHRs manager. The combination of hash checking and file verification, and user authentication of smart contract makes our system resistant with external attacks and significantly improves system integrity. 

- Additionally, our blockchain uses community validation to establish a decentralized healthcare network among cloud entities, healthcare users and smart contract, where all participants are synchronized by transaction ledgers. Any modifications caused by curious users are reflected on the blockchain and such malicious transactions are discarded from the network via consensus [2]. Users also share equal data management rights with the ability to monitor transactions, which in return guarantees data ownership and system reliability.

\section{Conclusions }
This paper proposes a novel cooperative architecture of data offloading and data sharing for healthcare by levering edge-cloud computing and Ethereum blockchain. We first propose a privacy-aware data offloading scheme where MDs can offload IoT health data to the edge server under system constraints. Then, a new data sharing is introduced by using blockchain and smart contract to enable secure data exchange among healthcare users. Specially, we develop a reliable access control mechanism associated with a decentralized IFPS storage design on cloud. Various experimental results demonstrate the significant advantages of the proposed offloading scheme over other baseline methods in terms of reduced time latency, energy consumption, and better memory usage. Moreover, the data sharing scheme can achieve efficient user authentication and significantly enhance data retrieval speeds while preventing malicious access to our healthcare system. System evaluations also prove that the operation cost of smart contract is low, and system security is guaranteed, showing the feasibility of our scheme for healthcare applications.


\begin{thebibliography}{00}

\bibitem{b1} S. M. Riazul Islam et al., "The Internet of Things for Health Care: A Comprehensive Survey," \textit{IEEE Access}, vol. 3, pp. 678-708, 2015.
		
\bibitem{b2} Guo, Hao, et al., "Attribute-based Multi-Signature and Encryption for EHR Management: A Blockchain-based Solution," in \textit{2020 IEEE International Conference on Blockchain and Cryptocurrency (ICBC)}, pp. 1-5, 2020. 

\bibitem{b3} Dinh C. Nguyen et al., "A mobile cloud based IoMT framework for automated health assessment and management," in \textit{41st Conference of the IEEE Engineering in Medicine \& Biology Society (EMBC)}, 2019.
\bibitem{b4} M. Asif-Ur-Rahman et al.,  "Towards a heterogeneous mist, fog, and cloud based framework for the internet of healthcare things," \textit{IEEE Internet of Things Journal}, vol. 6, pp. 4049-4062, 2018.
\bibitem{b5} R. Saha et al.,  "Privacy ensured e-healthcare for fog-enhanced IoT based applications," \textit{IEEE Access}, vol. 7, pp. 44 536-44 543, 2019.
\bibitem{b6} H. Wu et al., "Mobile healthcare systems with multi-cloud offloading," in \textit{IEEE Int. Conf. on Mobile Data Management}, vol. 2, 2013, pp. 188-193.
\bibitem{b7} Navaz  et al. "Towards an efficient and Energy-Aware mobile big health data architecture." \textit{Comput. methods and programs in bio.}, 2018.
\bibitem{b8} R. M. Abdelmoneem et al., "A cloud-fog based architecture for IoT applications dedicated to healthcare," in \textit{IEEE ICC}, 2019, pp. 1-6.
\bibitem{b9} D. Giri et al., "Sechealth: An efficient fog based sender initiated secure data transmission of healthcare sensors for e-medical system," in \textit{IEEE Global Communications Conference}, 2017, pp. 1-6.

\bibitem{b10}	M. Min et al., "Learning-based privacy-aware offloading for healthcare IoT with energy harvesting," \textit{IEEE Internet of Things Journal}, vol. 6, pp. 4307-4316, 2018.

\bibitem{b12} J. Liu et al.,  "BPDS: A blockchain based privacy-preserving data sharing for electronic medical records," in \textit{IEEE GLOBECOM}, 2018, pp. 1-6.
\bibitem{b13} H. Guo et al.,  "Access control for electronic health records with hybrid blockchain-edge architecture," in \textit{IEEE Int. Conf. on Blockchain}, 2019. 
\bibitem{b14} S. Wang al., "A blockchain-based framework for data sharing with fine-grained access control in decentralized storage systems," \textit{IEEE Access}, vol. 6, pp. 38437-38450, 2018.
\bibitem{b15} Dinh C. Nguyen et al., "Blockchain for secure EHRs sharing of mobile cloud based e-health systems," \textit{IEEE Access}, pp. 66792-66806, 2019.
\bibitem{b16} M. T. de Oliveira et al., "Towards a blockchain-based secure electronic medical record for healthcare applications," in \textit{IEEE ICC}, 2019. 	
\bibitem{b17}	Dinh C. Nguyen et al.,  "Privacy-preserved task offloading in mobile blockchain with deep reinforcement learning," \textit{IEEE Transactions on Network and Service Management}, 2020, in press. 


\bibitem{b18}	Sigwele, Tshiamo, et al. "Intelligent and energy efficient mobile smartphone gateway for healthcare smart devices based on 5G" in \textit{IEEE Global Communications Conference (GLOBECOM)}, 2018.. 

\bibitem{b19}	I. Elgendy et al.,  "An efficient and secured framework for mobile cloud computing," \textit{IEEE Transactions on Cloud Computing}, 2018.
\bibitem{b20} W.-T. Sung and Y.-C. Chiang, "Improved particle swarm optimization algorithm for android medical care IoT using modified parameters," \textit{Journal of medical systems}, vol. 36, no. 6, pp. 3755-3763, 2012.	

\bibitem{b21} AWS Lambda with Cloud Front Lambda Edge Services [Online]. Available:  https://aws.amazon.com/lambda/edge/.
 
\bibitem{b22} Firebase Performance Monitoring - Google  [Online]. Available: https://firebase.google.com/docs/perf-mon. 

	\end{thebibliography}
\end{document}